\documentclass[12pt]{article}

\usepackage{url}
\usepackage{scicite}
\usepackage{graphicx}
\usepackage{mathtools}
\usepackage{times}
\usepackage{enumitem}
\usepackage{lineno}

\newenvironment{sciabstract}{%
\begin{quote} \bf}
{\end{quote}}

\title{Detecting and mitigating bias in algorithms used to disseminate information in social networks}

\author
{Vedran Sekara,$^{1\ast\dagger}$ Ivan Dotu,$^{2\ast\dagger}$ Manuel Cebrian,$^{3,4}$ \\Esteban Moro,$^{5}$ and Manuel Garcia-Herranz$^{6\ast}$\\
\\
\normalsize{$^{1}$IT University of Copenhagen, 2300 Copenhagen, Denmark}\\
\normalsize{$^{2}$UNICEF/GIGA, New York, NY, USA}\\
\normalsize{$^{3}$Department of Statistics, Universidad Carlos III de Madrid, Madrid, Spain}\\
\normalsize{$^{4}$Center for Automation and Robotics, Spanish National Research Council, Madrid, Spain}\\
\normalsize{$^{5}$Network Science Institute, Northeastern University, Boston, MA, USA}\\
\normalsize{$^{6}$UNICEF, New York, NY, USA}\\
\\
\normalsize{$^\ast$To whom correspondence should be addressed; E-mail:  vsek@itu.dk, jdoturodriguez@unicef.org, and }\\
\normalsize{mherranz@unicef.org.}\\
\normalsize{$^\dagger$Contributed equally to the paper.}
}

\date{}

\begin{document} 


\baselineskip24pt


\maketitle 



\begin{sciabstract}
Social connections are conduits through which individuals communicate, information propagates, and diseases spread. Identifying individuals who are more likely to adopt ideas and spread them is essential in order to develop effective information campaigns, maximize the reach of resources, and fight epidemics. Influence maximization algorithms are used to identify sets of influencers. Based on extensive computer simulations on synthetic and ten diverse real-world social networks we show that seeding information using these methods creates information gaps. Our results show that these algorithms select influencers who do not disseminate information equitably, threatening to create an increasingly unequal society. To overcome this issue we devise a multi-objective algorithm which maximizes influence and information equity. Our results demonstrate it is possible to reduce vulnerability at a relatively low trade-off with respect to spread. This highlights that in our search for maximizing information we do not need to compromise on information equality.
\end{sciabstract}


\section*{Introduction}

Social relationships serve as important vectors through which a multitude of behaviors spread, from health related behaviors~\cite{centola2010spread,christakis2007spread}, innovation~\cite{rogers2010diffusion}, decisions of micro-financing~\cite{banerjee2013diffusion}, happiness~\cite{fowler2008dynamic}, cultural tastes~\cite{berger2009adoption}, to the emergence of social movements~\cite{gonzalez2011dynamics}.
Knowing through which pathways information spreads is vital for international development~\cite{barrett2014toward} and crucial for developing efficient methodologies that maximize the diffusion of potentially life saving information~\cite{chami2016profiling, alexander2022algorithms}.
Due to resource constraints it is unfeasible to send a piece of information to all individuals within a network.
Instead, a frequently adopted strategy is to seed a small set of individuals, much smaller than the full population, located at strategic places in the network whose activation (or removal) would facilitate the spread of information (or in the case of epidemics inhibit a disease from spreading). 
Numerous methods have been proposed to identify ``this set of influential nodes". 
The methodologies can be divided up into two fundamentally distinct classes~\cite{borgatti2006identifying,radicchi2017fundamental}, \textit{superspreader} and \textit{superblocker} methods. 
Superspreader methods identify individuals who are highly connected and effective at diffusing information~\cite{kempe2003maximizing,chen2009efficient,kitsak2010identification,banerjee2013diffusion,lokhov2017optimal,aral2018social}.
Superblocker methods identify individuals that occupy structurally vital positions in a network whose removal would destroy the network and subsequently block information from propagating~\cite{chen2008finding,morone2015influence,clusella2016immunization,zdeborova2016fast,braunstein2016network}.
Although the methods are different there is a general consensus that they both pinpoint nodes which are highly efficient conduits for information propagation~\cite{radicchi2017fundamental}.

Unfortunately, we have a limited understanding of which demographics these methods reach and which communities they leave behind, but there are alarming signs in the literature.
Previous work has shown that an individuals' chances of being ranked as an influencer are highly correlated with personal economic status~\cite{luo2017inferring}.
Similarly, influence maximization methods has been shown to create gaps in information access~\cite{fish2019gaps}, and to have a gender skew with male individuals having an advantage in being selected as influencers~\cite{stoica2019fairness,stoica2020seeding,jalali2020information}.
Further, social systems display strong levels of homophily, where connections between similar individuals occur at higher rates than between dissimilar individuals, with individuals being more likely to befriend people that resemble them~\cite{leo2016socioeconomic}.
As a consequence, information in social networks tends to be localized within social strata, restricting diffusion across demographic and socioeconomic gaps. 
In general, access to information is a major factor of social vulnerability~\cite{shirley2012social}.
By naively using current influence maximization methodologies to select influencers we run the risk of tailoring information campaigns towards the most affluent groups of our societies, while under-representing the most vulnerable and marginalized.


\subsection*{Defining Informational Vulnerability}
Vulnerability is a complex issue determined by physical, economic, social, and environmental factors, which decrease the capacity of individuals and groups to cope, anticipate, and react to hazards~\cite{turner2003framework}.
Here we study one aspect of vulnerability, namely individuals' access to information.
Previous work has looked into the interplay between influence maximization and characteristics of nodes (such as age, gender, ethnicity) which receive information~\cite{stoica2020seeding,ijcai2019p831,jalali2020information,wang2022information}.
Here we adopt a different approach. 
Rather than assuming we have access to demographic information of nodes we focus on the general utility of the information individuals receive---this is also called a `welfare approach'~\cite{heidari2018fairness,fish2019gaps}.
We study how access to information is distributed across all nodes in the network using numerical simulations.
Specifically, we define informational vulnerability as the average likelihood that an individual receives information, estimated from many independent diffusion cascades.
We focus on two aspects of this stochastic process: 1) \textit{frequency}: how often does an individual receive information (i.e. how often are they reached by cascades), and 2) \textit{recency}: how old are the cascades when they reach them  (i.e. at what step of a cascade is an individual reached). 
To simulate information diffusion processes in social networks we apply the commonly used Independent Cascade Model (ICM) in its most simple form, unweighted and undirected.
ICMs are commonly used to study influence maximization in social networks~\cite{kempe2003maximizing,beaman2021can}.
The ICM allows an informed individual one attempt to convince their neighbors to adopt a behavior (according to some probability); if successful the neighbors will try to convince their neighbors, etc. (see Supplementary Materials (SM) Sec.\ S3).

To identify sets of influencers we focus on two state-of-the-art influence maximization methods: degree discount~\cite{chen2009efficient} (DD) which is an efficient method for identifying superspreaders, and coreHD~\cite{zdeborova2016fast} (CHD) which infers superblockers. 
For comparison purposes we also include two commonly used heuristic for selecting influencers: highest degree~\cite{albert2000error} (HD), which selects nodes according to their number of connections, and k-core~\cite{kitsak2010identification} (KC) which selects nodes located in the core of the network.
Fig.\ 1a shows, for a small real-world network, which nodes each method selects as influencers.
As ICMs are stochastic we average over multiple realizations.
For each realization of the dynamic process we track which nodes are activated and how long it takes for the spreading process to reach them.
We quantify this using two measures.
The first measure, \textit{information frequency}:
$$\nu_i = \frac{1}{M} \: \sum_n^M a_{i,n} \: ,$$
summarizes the average fraction of times node $i$ has been reached, where $a_{i,n} = 1$ if node $i$ received information in realization $n$ and zero otherwise, and $M$ is the total number of realizations.
Information frequency ($\nu_i$) lies in the interval 0 to 1, where zero indicates that a node is never reached by any cascade, while a value of 1 indicates the node is reached by all cascades.

The second measure, \textit{information recency}:
$$\tau_i = \frac{1}{M} \: \sum_n^M \frac{1}{t_{i,n} + 1} \: ,$$
quantifies the temporal delay from process initialization ($t=0$) until node $i$ is reached.
Recency is calculated as the average of the inverse activation time to handle cases where the information spreading process dies out before reaching a node.
Nodes that on average receive information fast have $t_{i,n} \rightarrow 0$ and $\tau \rightarrow 1$, while nodes that are reached very late, or never, ($t_{i,n} \rightarrow \infty$) have $\tau \rightarrow 0$.

To uncover the shortcomings of the four influencer heuristics we compare them to a benchmark model where all nodes have an equal chance of being selected (selected at random) --- we call this the \textit{effective} measure.
If the ratio $\nu_i^{\text{method}}/\nu_i^{\text{benchmark}} > 1$ node $i$ will on average receive information more frequently when seeds are selected using a specific influencer mazimization heuristic as compared to when nodes are selected at random.
If $\nu_i^{\text{method}}/\nu_i^{\text{benchmark}} < 1$ node $i$ will be better off when information is inserted at random entry points in the network.
The same holds for recency, $\tau_i^{\text{method}}/\tau_i^{\text{benchmark}} > 1$ indicates node $i$, on average, receives more recent information when using a influencer maximization method, while $\tau_i^{\text{method}}/\tau_i^{\text{benchmark}} < 1$ denotes that information is received faster when information is inserted at random nodes.
Fig.\ 1b illustrates the resulting effective recency-values from using influencers inferred by the four heuristics as seeds.
Independent of influence maximization methodology, the influencer nodes and their surrounding neighbors are always reached by the influencer set, however, a large fraction of nodes seems to be left behind.
Typically nodes located on the periphery.

\begin{figure}[!htp]
\centering
\includegraphics[width=0.7\linewidth]{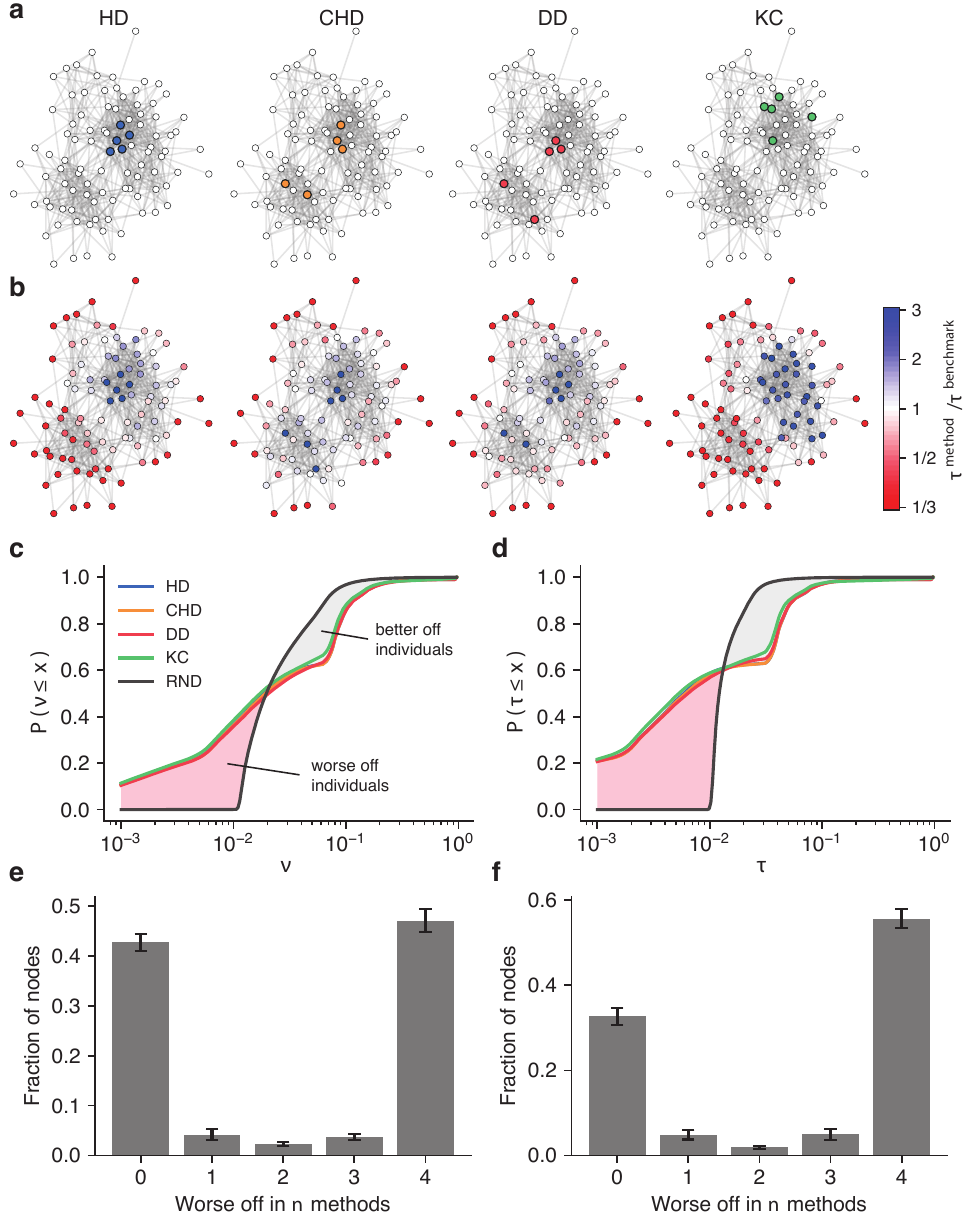}
\caption{\textbf{Information is unequally distributed in networks}. \small{\textbf{a}, Initial seed sets selected according to HD, CHD, DD, and KC, showing variations in how the four methods select influencers for a social network between households in a south-Indian village~\cite{banerjee2013diffusion}. Here, $5\%$ of nodes (colored) are selected as influencers for illustrative purposes (1\% otherwise). \textbf{b}, Effective recency for the social network. Recency is estimated across 1000 runs with infection probability $p = 0.069$ (see SM Sec. S3.1). \textbf{c}, Cumulative distribution of information frequency for synthetic SF-networks with $N = 10^4$, $\gamma = 2.5$, and average infection probability $\langle p \rangle = 0.085$. The curves show the probability that $\nu$ is less than or equal to $x$, where $x$ is any arbitrary value. Results are combined over 100 different network realizations. For each network we select 1 \% of nodes as influencers (inferred by one of the heuristics), run the spreading process, track which nodes receive information, and repeat the process $M = 10N$ times to account for stochasticity. Red shaded regions denote parts of the distribution where the effective measure is below one, while grey shaded indicate places where the ratio is above one. \textbf{d}, Cumulative distribution of recency for SF networks. \textbf{e}, Fraction of nodes that are worse off with respect to information frequency in $n$  of the seeding heuristics when compared to the benchmark. Error bars are standard deviation over 100 network realizations. \textbf{f}, Fraction of nodes that are worse off with respect to recency.}}
\label{fig:fig1}
\end{figure}

\subsection*{Quantifying Informational Vulnerability}
To formalize our observations from Fig.\ 1b we first investigate the four influencer maximization heuristics on a testbed of synthetic unweighted and undirected networks with scale-free (SF) degree distributions (see SM Sec.\ S6 for synthetic networks with normally distributed degree distributions).
While perfect SF networks are rarely observed in nature they are powerful simplifications of real-world networks~\cite{broido2019scale}.
In order to compare heuristics we construct influencer sets from a fixed finite fraction of the network population---1\% of nodes (see SM Sec.\ S5 for other seed sizes).

Seeding information through random nodes in SF networks results in a near-homogeneous frequency distribution (Fig.\ 1c black line).
In Fig.\ 1c, a completely equal distribution of information would be characterized by a vertical line in the cumulative probability distribution.
(Note that the black line characterizing the random process is not vertical due to the intrinsic variations of network structures.)
Using the four influence maximization methods to select influencers, however, results in fundamentally different frequency distributions (Fig.\ 1c, colored lines). 
Approximately half of nodes have effective frequency values above one, meaning they receive information more frequently then expected compared to a random process, while the other half receives much less (see also SM Sec.\ S4).
We observe similar results for recency, albeit slightly more polarized (Fig.\ 1d).
Looking across the different influencer heuristics, Fig.\ 1e, illustrates that if a node is under-informed by one heuristic, it will most likely not be better informed by any other heuristic.
On average $42.8 \pm 1.7 \%$ of nodes are better informed when information is seeded using any influencer maximization heuristics. We say that these nodes are always better off. However, $47.0 \pm 2.2 \%$ of nodes are consistently left behind, independent of which method is used to select influencers.
With respect to recency, Fig.\ 1f shows an even worse situation, $55.6 \pm 2.2 \%$ of nodes receive out-of-date information, independent on which information maximization methodology is used.


To understand the implications of information inequalities for real-world networks we look at 10 social networks, encompassing communication, interaction, and collaboration networks varying in size from hundreds to tens of thousands of individuals.
The networks are diverse in context, ranging from: face-to-face encounters~\cite{infect}, connections between households in multiple villages~\cite{banerjee2013diffusion}, connections between bloggers and blogs on political topics~\cite{adamic2005political}, digital communication between university students~\cite{opsahl2009clustering}, email communication~\cite{guimera2003self}, online friendships on Facebook~\cite{cho2011friendship}, and scientific collaborations~\cite{leskovec2007graph,newman2001structure} (see SM Sec.\ S2 for details).

Repeating the analysis from Fig 1 for these real-world networks, we find influencer maximization heuristics to, on average, leave significant portions of the networks in disadvantaged positions in terms of information frequency (Fig.\ 2a).
Overall HD, CHD, and DD result in fairly similar $\nu$ distributions, while selecting influencers according to KC performs worse (with up to 80\% of nodes being worse off). 
This is due to real-world network having a large numbers of cliques, where small discrepancies in shell numbers can result in KC only selecting nodes from a single clique~\cite{kitsak2010identification}, effectively limiting the diffusion of information.
Similar behavior is observed for information recency.
Fig.\ 2b shows that HD, CHD, and DD leave behind a comparable numbers of nodes, while KC consistently performs worse.
Summarizing the average reach of influencer heuristics, we find that up to $69.8\%$ of a network might receive information less frequently compared to if it is input at random (Fig.\ 2c), and information can reach up to $79.2\%$ of individuals slower (Fig.\ 2d).

There is a connection between access to information and a node's position in the network.
The connection is so strong that a predictive model, based on structural features of as node can accurately predict whether a node will fall into the group of `worse off according to all influencer heuristics` or `better off in all' (SM Sec.\ S8). 
We can, on average, with 97.4\% accuracy predict the information status of nodes regarding frequency, and 96.9\% for recency.
This demonstrates that current influencer heuristics suffer from biases which disadvantage low-connected and peripheral nodes.

\begin{figure}[!htp]
\centering
\includegraphics[width=0.99\linewidth]{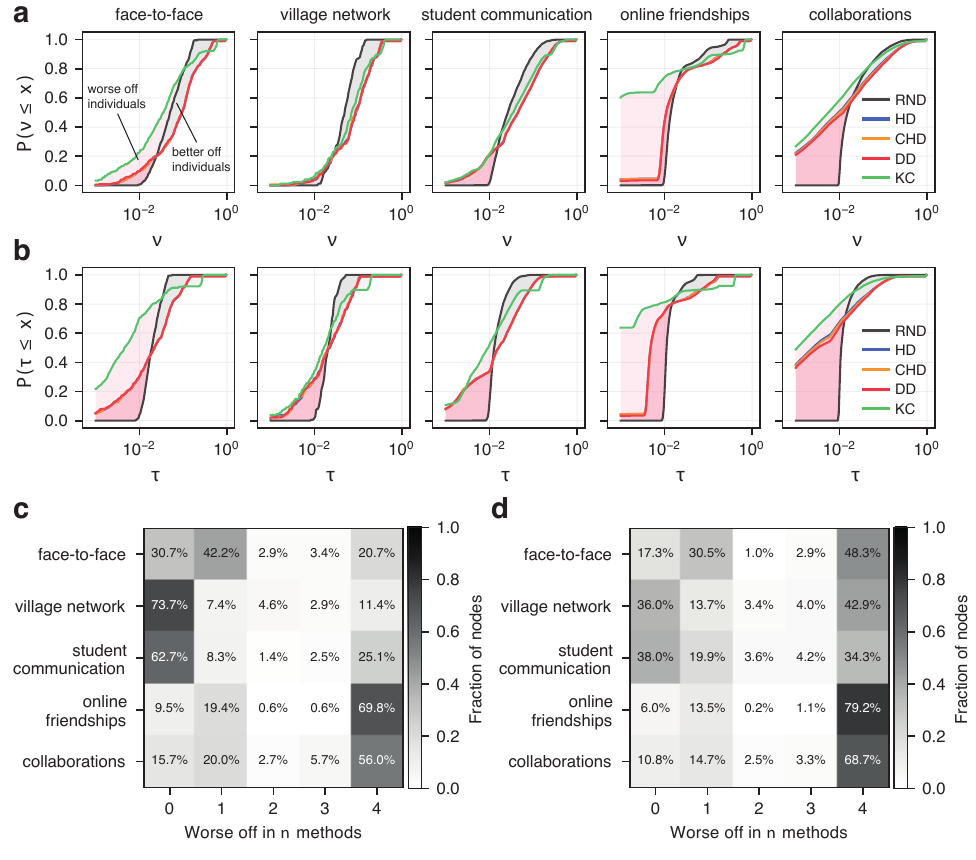}
\caption{\textbf{Information is unequally distributed in real-world social networks}. Here we show results for five of the networks, see SM Fig S9 for results for other networks. \textbf{a}, Cumulative distribution of individual node frequency for networks ordered according to size (number of nodes). Initial seeds contain $1\%$ of network nodes, and results are  averaged over $10N$ simulations (see SM Table S1). \textbf{b}, Cumulative distribution of recency for empirical networks. \textbf{c}, Fraction of nodes that are worse off with respect to frequency in $n$ of the seeding heuristics when compared to the random seeding procedure ($\nu^{\text{method}}/\nu^{\text{benchmark}}$). Demonstrating that large parts of social networks are in disadvantaged positions. \textbf{d}, Fraction of nodes that are worse off with respect to recency ($\tau^{\text{method}}/\tau^{\text{benchmark}}$) for $n$ methods.}
\label{fig:fig2}
\end{figure}


\subsection*{Fair Influence Maximization}
Different strategies can be employed to bridge the information gap. 
Previous work have shown that instead of using influencer algorithms to identify $s$ individuals, one can select slightly more individuals $s+\epsilon$, but at random~\cite{akbarpour2018diffusion}.
Even for small $\epsilon$-values this has been shown to result in larger cascades.
Another solution is to apply acquaintance methods, also called friendship-nomination, which work by selecting a random neighbor of a randomly selected node. These have been used for mass drug administration campaigns~\cite{chami2017social}, to seed information about maternal and child health~\cite{airoldi2024induction}, and for inferring centrally located individuals suitable as monitors for detecting large scale disease outbreaks~\cite{garcia2014using}.
Lastly, algorithms that use iterative realizations of ICMs to equitably maximize social welfare objective functions have been proposed to bridge the information gap~\cite{fish2019gaps}, however, we find them not to be effective at bridging the information gap (see SM Fig.\ S12).

We embark on a different solution.
Traditionally, influence maximization has only focused on maximizing a single objective function, information spread.
However, as literature from the field of Artificial Intelligence shows, focusing only on one parameter can lead to troubling and unfair outcomes~\cite{obermeyer2019dissecting,thomas2022reliance}. 
Instead, we propose a multi-objective formulation of the fair influence maximization, where both spread and fairness are taken into account in the fitness of a candidate solution (set of selected influencers).
We measure fairness as the fraction of nodes that receive information at a higher frequency and speed than what is expected from the benchmark model.
We say a node is vulnerable if it receive less information than what can be expected at random ($\nu_i^{\text{method}}/\nu_i^{\text{benchmark}} < 1$).
In the following we focus primarily on information frequency, but similar results can be obtained for recency.
The more fair an influencer set is, the fewer nodes will be vulnerable, so we maximize the number of non-vulnerable nodes in addition to maximizing spread. 
Analytically calculating how information will spread from a set of nodes according to the ICM model is a computationally hard problem (NP-hard)~\cite{kempe2003maximizing}.
Instead, we use an approximation of the fitness of a influencer set (see SM Sec.\ S9).
To find fairer seeds we then use a genetic algorithm to solve the optimization problem (see Methods).

\begin{figure}[!htp]
\centering
\includegraphics[width=0.8\linewidth]{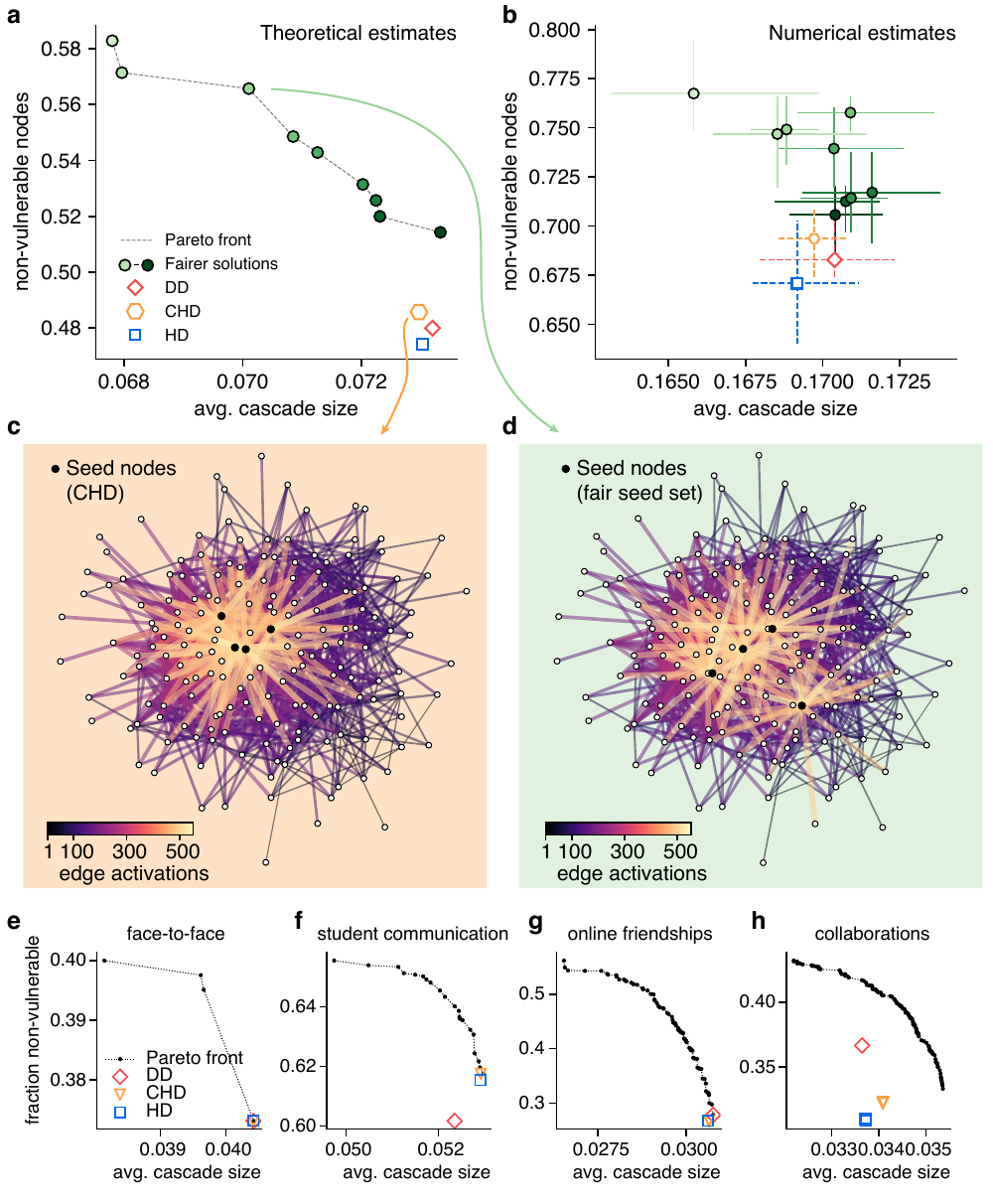}
\caption{\textbf{Fair influence maximization for social networks}. \textbf{a}, Theoretical Pareto front of optimal influencer sets identified by our multi-objective algorithm for the social network between households in a South Indian village, compared to influence maximization heuristics. We disregard KC as it consistently performs worse, both in terms of fairness and reach compared to the other heuristics. Higher values of non-vulnerable nodes indicate higher values of fairness. \textbf{b}, Numerical evaluation of influencer sets using ICMs. Error bars are given as the standard deviation from 10 realization of $10N$ ICM simulations. \textbf{c}, Edge activations for the set of influencers identified by CHD. Edges are colored and sized according to how often they are activated during $10N$ simulations. Nodes colored black are seed nodes. \textbf{d}, Edge activations for one of the fairer seed sets identified by our algorithm. A comprehensive comparison between the seed sets is available in SM Sec. S10. \textbf{e-h}, Theoretical Pareto fronts for four additional real-world social networks (see SM Fig.\ S15 for results for remaining five networks, and for numerical results from ICMs).}
\label{fig:fig3}
\end{figure}

Fig.\ \ref{fig:fig3}a shows the theoretical Pareto front (in multi objective optimization a Pareto front denotes a line of optimal solutions) for a social network between households in a village.
Our method identifies seed sets which are more fair and, at the same time, as effective at maximizing influence as the influence maximization heuristics (HD, CHD, DD).
(We disregard here seed sets identified by KC as they are far inferior to the ones produced by the other methods.) 
For this network, our algorithm identifies nine possible influencer sets, undiscovered by the traditional heuristics, each with different trade-offs between maximizing information reach (cascade size) and the number of non-vulnerable nodes (fairness).
As these findings are based on an theoretical approximation we also evaluate them numerically using ICMs.
Fig.\ \ref{fig:fig3}b, shows our theoretical predictions are consistent with results from numerical simulations.
For a negligible reduction in cascade size we can, for this specific network, choose fairer seeds that roughly corresponds to 6\% to 10$\%$ fewer vulnerable nodes. 
Fig.\ \ref{fig:fig3}c-d illustrates difference between seed sets inferred by a state-of-the-art influencer maximization heuristics (CHD) and our approach. 
The figure shows initial seed nodes (in black) and the activation of edges, where an activated edge indicates successful information propagation.
Visually there is a clear difference between which nodes are selected as influencers.
Using the average distance from seed nodes to the rest of the network we calculate how far each seed set is to all nodes in the network. We find that our method identifies nodes which are more evenly distributed in the network and, on average, closer to the overall network (SM Sec.\ S11).
This results in larger parts of the network being more easily reached.

Fig.\ \ref{fig:fig3}a,b results for one network, however, our fair influence maximization method works equally well for other real-world networks (Fig.\ \ref{fig:fig3}e-h, and SM Fig. S15).
For other networks (Fig.\ \ref{fig:fig3}e-h) we find similar results. 
While influence maximization heuristics identify seed sets which optimize cascade size, we find that these seed sets are not fair in terms of information equality (for numerical results see SM Sec. S10).
For all networks it is possible to improve this, with large gains in fairness being achievable with small trade-offs in cascade size. 
Overall we find that a $1\%$ decrease in cascade size can result in a decrease in number of vulnerable individual by approximately: $1.9\%$ for the network of political blogs, $3.3\%$ for the email communication network, $3.4\%$ for student communication, up to $10.6\%$ for collaboration networks, and $24\%$ for online friendships on Facebook (here we disregard the face-to-face and village networks due to the low number of identified seed sets).

Lastly, for the collaboration networks (Fig.\ \ref{fig:fig3}h), we find that seed sets identified by influence maximization heuristics are not even close to the Pareto frontier.
Meaning they are sub-optimal both in term of cascades sizes and fair information access.
In this situation our algorithm can also be used to identify, previously undiscovered, seed sets which optimize cascade-size.

\subsection*{Discussion}
The United Nations sustainable development goals (SDGs) recognize that eradicating inequalities in all their forms and dimensions are one of the greatest global challenges our societies face. 
Algorithms have the power to deliver on the SDGs.
For example, access to information is critical for vaccination campaigns and algorithms such as influence maximization have a role to play in effectivizing these campaigns.
However, algorithms can also bring potential biases into play.
Our results show there are groups of nodes that are consistently left behind by influence maximization algorithms.
In particular, for both real-world and synthetic networks, we find that access to information in unequal, both in terms of how often information is received and how recent the information is.
A behavior that is not limited to low or high clustered networks, nor to specific types of interactions (SM Table S1); we find it present across all networks we investigated.

Although algorithmic systems can be biased due to many factors~\cite{mehrabi2021survey}, it is often thought that biases appear due to skews or misrepresentations in training data.
However, that is not the case for influence maximization algorithms.
Here, the issue lies with the problem statement and the choice of objective function.
An algorithmic bias is created by focusing algorithms solely on optimizing reach, without considering information equity.
Unfortunately, not receiving information has real-world consequences. 
For example, experiences from mass drug administration campaigns have pinpointed that individuals are left untreated not due to lack of medicine, but because they never receive information about the campaign~\cite{chami2016profiling}.
The pervasive usage of influence maximization algorithms in information diffusion and in online social network~\cite{leskovec2007cost,coro2021link} can create large fractures in the social fabric of our societies.

Thus, it is vital to understand if such algorithms are equitable, to quantify the level of inequality, and propose potential alternatives that can balance potential reach and equity.
Multi-objective-optimization is a well-known computational tool which adds nuance to optimization problems and enables the inclusion of multiple criteria. 
Our results demonstrate it is possible to find influencer sets that reduce vulnerability at a relatively low trade-off with respect to spread.
For example, we find that a mere $1\%$ reduction in reach can reduce the number of people left behind in information campaigns by up to $24\%$.
Further, we find that our approach is more effective at bridging the information gap compared to other proposed solutions, for example the Myopic method (see SM Sec.\ S9.1).

Our multi-objective algorithm is a first approach at solving this critical problem, yet it is not perfect.
We believe it can act as a starting point towards more systematic solutions towards fair information access, as this issue arise across many other contexts within network science, artificial intelligence, and computational science problems.
One particular application can be online social networks where incorporating additional algorithmic objectives can be beneficial to: help detect vulnerable individuals, mitigate and reduce segregation, lessen polarization between groups, and help guide the design of more equal information dissemination structures. 

Our approach requires information about the full network. 
Noise and incomplete mappings of networks will naturally affect this.
However, we believe the effect will not differ from what methods like degree discount (DD), CoreHD (CHD) or highest degree (HD) already experience, as they also require information about the total graph.
Another shortcoming is that we focus purely on simple contagion effects, where nodes have equal, and independent probabilities of adopting a behavior. 
Complex contagion, where individuals require social affirmation from multiple sources, has been observed for certain social settings, including sharing of content on social media~\cite{romero2011differences,monsted2017evidence}, and online behaviors~\cite{centola2010spread}.
The nature of contagion depends on the type of the situation, and whether interactions happen at a local or global level.
We focus on simple contagion because it is believed to be the main factor in information spreading~\cite{centola2007complex}.
For instance, if a person is looking for job opportunities, it is more beneficial to receive information from the global network, via weak ties, rather than just from close friends and family~\cite{granovetter1973strength,rajkumar2022causal}.
However, future work should understand how information inequalities develop in complex contagion scenarios.
Further, our study focuses on static and undirected networks, but information propagation in real world often occurs through directed and temporal connections. Extending the framework to such networks and investigating if similar inequalities emerge is an equally important direction for future work.

Lastly, our definition of information inequality relies on benchmarking existing methods to random information spreading scenarios, as this is the most \textit{`fair'} system we can imagine.
Other definitions can also be used, and future work should focus on testing them. 
Independent of the choice of definition, it is vital that inequalities, which arise, or are amplified, as result of algorithms, be quantified and measured.
As our world is becoming increasingly digitalized, access to correct, timely, and factual information will grow in significance.
As such, it is critical to know how well algorithms which deal with information dissemination and delivery work, and which groups and individuals they leave behind.



\bibliography{sample}

\bibliographystyle{Science}

\section*{Acknowledgments}
MGH and ID want to thank AECID (Spanish Agency for International Development Cooperation) for their support to data innovation and Frontier Data Technologies through UNICEF’s Frontier Data Network. MC wishes to acknowledge the following funding: Ministerio de Ciencia, Innovación y Universidades” y a la "Convocatoria de la Universidad Carlos III de Madrid de Ayudas para la recualificación del sistema universitario español para 2021-2023, de 1 de julio de 2021" en base al Real Decreto 289/2021, de 20 de abril de 2021 por el que se regula la concesión directa de subvenciones a universidades públicas para la recualificación del sistema universitario español". Furthermore, MC is thankful for the support of project “Ayuda PID2022-137243OB-I00 financiada por MCIN/AEI/10.13039/501100011033” and by “FEDER Una manera de hacer Europa”. EM acknowledges support by the National Sience Foundation under grant No. 2218748.


\section*{Methods}
\subsubsection*{Independent cascade model}
The ICM process is as follows: at time $t=0$ all nodes are inactive, except for initial seed nodes (activated).
At each time step $t$ an activated node $i$ will contacts all its neighbors, which have previously not been activated, and try to convince/activate them according to an independent transmission probability $p$.
After attempting to convince all its neighbors a node becomes inactive and cannot be activated again in subsequent stages of the dynamic.
The process is iterated until no active nodes remain.

\subsubsection*{Infection probability}
For ICM the only parameter is the activation probability $p$ (probability of convincing people to adopt a behavior); a too high probability will correspond to a global information cascade with the full network adopting a behavior, a too low would entail no information spreading.
We set $p = p_c$ where $p_c$ is the critical probability separating the region of the phase diagram where cascades (outbreaks) are subextensive ($p < p_c$) from the supercritical region ($p > p_c$) where outbreaks reach a finite fraction of the whole network~\cite{radicchi2017fundamental}.
For each network we calculate the critical value of the transmission probability ($p_c$) as the position of the maximum of the susceptibility $\langle s^2\rangle/\langle s \rangle^2$, where $\langle s^n\rangle$ is the n-th moment of the outbreak size distribution computed for random selected initial single spreaders~\cite{castellano2016numerical}.
See SM for more information.

\subsubsection*{Fair Influence Maximization}
We implement a simple version of a non-dominated sorting genetic algorithm (NSGAIII) using the `distributed evolutionary algorithms in python' (DEAP) library~\cite{DEAP_JMLR2012}. The code is freely available at [link will be added when manuscript is accepted]. Briefly, the main modelling set-up is:
\begin{itemize}
    \item \textbf{Initialization} is performed by generating one set of influencers with each of the heuristics mentioned in this paper and the rest completely at random.
    \item \textbf{Crossover} of two individual sets is performed by generating the union of both sets and choosing from that sets the seeds for both new individuals at random.
    \item \textbf{Mutation} is composed of two operators that are performed with different frequencies:
    \begin{enumerate}[label=(\alph*)]
        \item \textbf{Random} where $10\%$ of the seeds are removed from a individual set and new ones selected at random.
        \item \textbf{Tabu-like} where one seed is removed at random and a certain number (or all) of random seeds are inspected for addition to the set, and the one with lowest vulnerability is finally selected.
    \end{enumerate}
    
\end{itemize}

\noindent All experiments on empirical networks were run with a population of $100$ individual sets, a mximum of $100$ generations, crossover probability of $0.8$, mutation probability of $1$, Tabu-like mutation frequency of $0.4$ and size of the tabu neighborhood of $20\%$ of the total number of nodes on the network.

\end{document}